\documentclass{ws-procs975x65}
\newcommand{\bea}{\begin{eqnarray}}
\newcommand{\eea}{\end{eqnarray}}
\newcommand{\be}{\begin{equation}}
\newcommand{\ee}{\end{equation}}
\newcommand{\beast}{\begin{eqnarray*}}
\newcommand{\eeast}{\end{eqnarray*}}
\newcommand{\pkt}{\; .}
\newcommand{\kma}{\; ,}
\newcommand{\nn}{\nonumber}
\def\e{{\rm e}}

\usepackage[]{graphicx}
\usepackage[]{subfigure}

\newcommand{\fullWidth}{390pt}

\begin{document}

\title{Wormhole creation by quantum tunnelling}

\author{Lorenzo Battarra$^a$,~~~George Lavrelashvili$^{a, b}$~~~and~~~Jean-Luc Lehners$^a$}
%%%Andrew B. Author$^*$ and Charles D. Author

\address{$^a$Max-Planck-Institute for Gravitational Physics \\
Albert-Einstein-Institute, D-14476 Potsdam, Germany \\
$^b$Department of Theoretical Physics,
A.Razmadze Mathematical Institute \\
I.Javakhishvili Tbilisi State University,
GE-0177 Tbilisi, Georgia}

\begin{abstract}
We study the process of quantum tunnelling in self-interacting scalar field theories with non-minimal coupling to gravity.
In these theories gravitational instantons can develop a neck -- a feature prohibited in theories with minimal coupling, and describing the nucleation of geometries containing a wormhole. We also clarify the relationship of neck geometries to violations of the null energy condition.
\end{abstract}
\keywords{Wormholes; Quantum tunnelling; Phase transitions with gravity.}
\bodymatter
%%%%%%%%%%%%%%%%% now a standard article style for the most part
%%%%%%%%%%%%%%%%%%%%%%%%%%%%%%%%%%%%%%%%
\section{Introduction} \label{intro}
%%%%%%%%%%%%%%%%%%%%%%%%%%%%%%%%%%%%%%%%
Recently there has been substantial interest in theories violating the null energy condition (NEC)
(see e.g. \cite{Rubakov:2014jja}, \cite{Elder:2013gya} for reviews).
Such theories may lead to interesting phenomena like the creation of a universe in the laboratory \cite{Rubakov:2014jja}, the existence of traversable Lorentzian wormholes \cite{visser95} or non-singular bounce solutions \cite{Buchbinder:2007ad,Creminelli:2007aq,Easson:2011zy,Koehn:2013upa,Battarra:2014tga,Koehn:2015vvy}.
One of the examples of NEC violating theories is a scalar field theory non-minimally coupled to gravity \cite{Flanagan:1996gw}
and Lorentzian wormholes were in fact found in this theory \cite{Barcelo:1999hq,Barcelo:2000zf}.
Lorentzian wormholes typically join two asymptotically flat geometries, or could be a bridge between
an asymptotically flat and a spatially closed universe.
The characteristic feature of a wormhole is the existence of a ``neck'' in a spatial slice.

In the present study we will consider the Euclidean version of modified gravity theories
and will be interested in the possibility of creating a wormhole
during metastable vacuum decay processes \cite{Coleman:1980aw}.
A priori there are four possible instanton shapes in de Sitter to de Sitter transitions,
depending on whether the false and true vacuum regions are smaller or larger than half of Euclidean de Sitter space.
A neck can only be present in the case where both ``halves'' of the instanton are larger than half of Euclidean de Sitter space.
However, it was shown in \cite{gl87} that in scalar field theories minimally coupled to gravity such configurations cannot arise.
At the same time it was argued  \cite{gl87} that the creation of instantons with necks might be possible if
one allows for a non-minimal coupling of the scalar field to gravity.
Here we will explore this possibility in detail\footnote{The present contribution is based on \cite{Battarra:2014naa}.}.
%%%%%%%%%%%%%%%%%%%%%%%%%%%%%%%%%%%%%%%%%%%%%%%%%%%%%%%%%%%%%%%%%%%%%%%%%%
\section{Minimal coupling and NEC violation} \label{minimal}
We will start with a simple model of a scalar field $\phi$ with a potential $V(\phi)$
minimally coupled to gravity and we will consider homogeneous and isotropic universes, described by the metric
%%%
\be
ds^2=-dt^2+a^2(t)\gamma^K_{ij} dx^i dx^j = -dt^2 + a^2(t)[\frac{dr^2}{1-K r^2}+r^2 d\theta^2 + r^2 \sin^2 (\theta){d\varphi}^2].
\ee
%%%
In what follows we will only be interested in the $K=+1$ case, but for clarity we will write $K$ out
explicitly in this section. The energy momentum tensor is given by
$T_{00}=\rho_s \; , \; T_{ij}=a^2 \gamma^K_{ij} p_s$,
where the energy density and the pressure are given respectively by
%%%
\be \label{energypressure}
\rho_s =\frac{1}{2} \left({\frac{d\phi}{dt}}\right)^2 + V \; , \quad \; p_s = \frac{1}{2} \left({\frac{d\phi}{dt}}\right)^2 - V  \pkt
\ee
%%%
The null energy condition (NEC) $T_{\mu\nu}n^\mu n^\nu > 0$,
with $n_\mu$ being a null vector, $n_\mu n^\mu=0$, then reduces to the requirement
%%%
\be \label{NEC}
\rho_s+ p_s > 0 \pkt
\ee
%%%
The equations of motion (Friedmann equations) can be written in the form
%%%
\be
H^2 = \frac{\kappa}{3}\rho_s -\frac{K}{a^2} \kma \quad
\frac{dH}{dt} = -\frac{\kappa}{2} (\rho_s + p_s) +\frac{K}{a^2} \kma
\ee
%%%
where $H \equiv (da/dt)/a$ and $\kappa$ is the reduced Newton's constant. 
Tunnelling can be described by performing an analytic continuation to Euclidean time, with $t=-i \bar{\lambda}$.
Then the metric and scalar field are of the form
$ d\bar{s}^2_E=d\bar{\lambda}^2 +\bar{\rho}^2 d\Omega^2_3 \kma \; \bar\phi=\bar\phi(\bar{\lambda})$,
where $\bar{\rho}(\bar{\lambda})\equiv a(it)$.
Note that the Euclidean version of the NEC condition Eq.~(\ref{NEC}) reverses sign:
%%%
\be \label{ENEC}
\rho^E_s+p^E_s < 0 \kma
\ee
%%%
where the Euclidean energy density and pressure are obtained by analytic continuation of Eq. (\ref{energypressure}),
$\rho^E_s = - \frac{1}{2} \left({\frac{d\phi}{d\bar{\lambda}}}\right)^2 + V \; ,  \; 
p^E_s = -  \frac{1}{2} \left({\frac{d\phi}{d\bar{\lambda}}}\right)^2 - V$.
The Euclidean versions of the Friedmann equations read
$H_E^2 = -\frac{\kappa}{3}\rho^E_s +\frac{K}{a^2} \kma \;
\frac{dH_E}{d\bar{\lambda}} = \frac{\kappa}{2} (\rho^E_s+p^E_s) -\frac{K}{a^2}$,
where $H_E = (d\bar{\rho}/d\bar{\lambda})/\bar{\rho}$. 
At the putative neck of an instanton, i.e. at a local minimum of $\bar\rho(\bar{\lambda}),$ we have
$H_E = 0$ and would need $ \frac{dH_E}{d\bar{\lambda}}>0$,
which, in view of the second (Euclidean) Friedmann equation, 
is impossible if the ``NEC'' condition Eq.~(\ref{ENEC}) is fulfilled. 
Thus we can see that ($O(4)$ symmetric) instantons in theories whose Lorentzian 
counterpart satisfies the NEC cannot have a neck.

%%%%%%%%%%%%%%%%%%%%%%%%%%%%%%%%%%%%%%%
\section{Modified gravity: Einstein and Jordan frames}
%%%%%%%%%%%%%%%%%%%%%%%%%%%%%%%%%%%%%%%

The arguments of the previous section motivate us to study theories in which the scalar field is non-minimally coupled to gravity. In particular, we will be interested in the theory defined by the Euclidean action
\begin{equation} \label{eq:jordan}
S_E = \int d ^4x \sqrt{g} \left( - \frac{1}{2 \kappa} f( \phi) R + \frac{1}{2} \nabla_\mu \phi \nabla^\mu \phi + V( \phi) \right)
+S_m (\psi_m, g_{\mu\nu}) \;,
\end{equation}
where the matter action $S_m$ depends on matter fields $\psi_m,$ which we assume to couple to the physical metric $g_{\mu\nu}$ \cite{Steinhardt:1994vs}.
With the conformal transformation and field redefinition
%%%
\be  \label{eq:phibphi}
g _{\mu\nu}  \equiv  f ^{-1}\, \bar{g} _{\mu\nu} \kma \quad
\frac{d \bar{ \phi}}{d \phi}  \equiv  \frac{\sqrt{ f + \frac{3}{2 \kappa} f_{, \phi} ^2}}{ f} \kma
\ee
%%%
we obtain the action in Einstein frame,
\begin{equation}  \label{eq:einstein}
S_E = \int d ^4x\, \sqrt{\bar{g}} \left(- \frac{1}{2\kappa}\bar{R} + \frac{1}{2} \bar{\nabla}_\mu \bar{\phi} \bar{\nabla}^\mu \bar{\phi} + \bar{V} \right)
+S_m (\psi_m, f^{-1} \bar{g}_{\mu\nu}) \;,
\end{equation}
where $\bar{V}=V(\phi(\bar{\phi}))/f^2$.
At the level of classical solutions, this means that if
$ds ^2  =  d \lambda ^2 + \rho ^2( \lambda) d \Omega _3 ^2 \kma \;  \phi  =  \phi( \lambda)$,
is a solution in Jordan frame (\ref{eq:jordan}), then
$ds ^2  =  d \bar{ \lambda} ^2 + f( \phi( \bar{ \lambda})) \rho ^2( \bar{ \lambda}) d \Omega _3 ^2 \kma \;
\bar{ \phi}  =  \bar{ \phi}( \phi ( \bar{ \lambda}))$,
is a solution in Einstein frame (\ref{eq:einstein}) provided that $ \bar{ \phi}( \phi)$ is specified (up to an irrelevant integration constant) by (\ref{eq:phibphi}) and $ \frac{ d \bar{ \lambda}}{d \lambda} = f ^{1/2} \pkt $
In particular, this means that the two ``scale factors'' are related by
\begin{equation} \label{eq:rhotransf}
\rho = \frac{ \bar{ \rho}}{ f ^{1/2}} \;.
\end{equation}
This implies that, if $ \bar{ \rho}$ is a ``normal'' instanton with only one extremum (local maximum)
and the function $f$ has a sufficiently sharp local maximum, the profile of the instanton in the Jordan frame can develop a neck.
%%%
\section{Non-minimal coupling: model and field equations} \label{model}
%%%%%%%%%%%%%%%%%%%%%%%%%%%%%%%%%%%%%%%%
%%%
For specificity we will choose 
$ f(\phi)=1-\kappa\xi \phi^2\; ,$
i.e. we will consider the Euclidean theory with action
%%%
\be \label{ea1}
S_E=\int d^4x \sqrt{g} \Bigl( -\frac{1}{2\kappa} R+ \frac{1}{2} \nabla_\mu \phi \nabla^\mu \phi
+V(\phi) +\frac{\xi}{2} \phi^2 R \Bigr) \kma
\ee
%%%
where $\xi$ is dimensionless parameter. Varying this action w.r.t. $\phi$ and the metric leads to the scalar field equation
%%%
\be \label{sc_eq1}
\nabla_\mu \nabla^\mu \phi -\xi R \phi = \frac{dV}{d\phi} \kma
\ee
%%%
and the gravity equations
%%%
\be\label{gr_eq1}
R_{\mu\nu}- \frac{1}{2} g_{\mu\nu}R = \tilde\kappa T_{\mu\nu}
- \tilde\kappa \xi (\nabla_\mu \nabla_\nu - g_{\mu \nu} \nabla_\lambda \nabla^\lambda) \phi^2 \kma
\ee
%%%
where
$\tilde{\kappa}\equiv \frac{\kappa}{1-\kappa \xi \phi^2}$
is the effective gravitational constant and the minimally coupled energy momentum tensor is given by
$T_{\mu\nu}=\nabla_\mu \phi \nabla_\nu \phi -\frac{1}{2} g_{\mu\nu} \nabla_\lambda \phi \nabla^\lambda \phi
-g_{\mu\nu} V (\phi) \pkt$
Assuming $O(4)-$symmetry, and introducing the notation ${\dot{}} \equiv d/d\lambda$, 
the equations of motion can be rewritten 
in a form that is convenient for numerical integration:
%%%
\bea
\label{phi_dd2}
\ddot{\phi}+3\frac{\dot{\rho}}{\rho} \dot{\phi}
- \frac{\kappa\xi\phi}{1-\kappa\xi (1-6 \xi) \phi^2} [4 V - 6\xi \phi \frac{dV}{d\phi}+(1-6\xi) \dot{\phi}^2]
= \frac{dV}{d\phi} \kma
\eea
%%%
\bea \label{rho_2}
\ddot{\rho} =-\frac{\tilde{\kappa} \rho}{3} \Bigl( (1-\frac{3\xi}{1-\kappa\xi (1- 6 \xi) \phi^2})\dot{\phi}^2
+\frac{1-\kappa\xi (1+6 \xi) \phi^2}{1-\kappa\xi (1-6 \xi)\phi^2} V  \nn \\
+ 6 \xi \frac{\dot{\rho}}{\rho} \phi \dot{\phi}
-\frac{3\xi (1-\kappa \xi \phi^2)}{1-\kappa \xi (1- 6 \xi)\phi^2} \phi \frac{dV}{d\phi}
\Bigr) \pkt
\eea
%%%
%Eqs.~(\ref{phi_dd2}), (\ref{rho_2}) simplify for the particular value $\xi=1/6$, which reflects
%the value for a conformally invariant coupling of a massless scalar field \cite{birrell}.
Note that the rhs of Eqs.~(\ref{gr_eq1}) allow us to find the Euclidean energy density and pressure for a non-minimally
coupled scalar field as
%%%
\begin{eqnarray}
\rho^E_{\xi} &=& \tilde\kappa \left( - \frac{1}{2} \left({\frac{d\phi}{d{\lambda}}}\right)^2 + V - 3 \xi H_E {\frac{d(\phi^2)}{d{\lambda}}}\right) \kma \\
p^E_{\xi} &=& \tilde\kappa \left( - \frac{1}{2} \left({\frac{d\phi}{d{\lambda}}}\right)^2 - V + \xi {\frac{d^2(\phi^2)}{d{\lambda}^2}}+ 2 \xi H_E {\frac{d(\phi^2)}{d{\lambda}}}\right) \pkt
\end{eqnarray}
%%%
Thus we see that now the Euclidean NEC,
%%%
$
\rho^E_{\xi} + p^E_{\xi} < 0 \quad \leftrightarrow \quad - \left({\frac{d\phi}{d{\lambda}}}\right)^2 + \xi {\frac{d^2(\phi^2)}{d{\lambda}^2}} - \xi H_E {\frac{d(\phi^2)}{d{\lambda}}} < 0 \,,
$
%%%
has the possibility of being violated if $\xi \neq 0$. Such violations due to non-minimal coupling were previously
discussed e.g. in \cite{Flanagan:1996gw,Visser:1999de,Barcelo:2000zf}.

We will now assume that the potential $V(\phi)$ is positive and has two non-degenerate local minima
at $\phi=\phi_{\rm tv}$ and $\phi=\phi_{\rm fv}$, with $V(\phi_{\rm fv})>V(\phi_{\rm tv})$,
as well as a local maximum for some $\phi=\phi_{\rm top}$, with $\phi_{\rm fv}<\phi_{\rm top}<\phi_{\rm tv}$.
The Euclidean solution describing vacuum decay at $\lambda=0$  satisfies the conditions
%%%
\be\label{eq:bcs}
\phi (0)= \phi_0,\qquad \dot{\phi}(0) = 0,\qquad \rho(0)=0, \qquad \dot{\rho}(0)=1 \kma
\ee
%%%
and similar behaviour at the other end of the instanton, at some $\lambda=\lambda_{max}$\cite{Battarra:2014naa}.
%%%%%%%%%%%%%%%%%%%%%%%%%%%%%%%%%%%%%%%%%%%%%%%%%%%%%%%
\section{Numerical Examples}
%%%%%%%%%%%%%%%%%%%%%%%%%%%%%%%%%%%%%%%%%%%%%%%%%%%%%%%
For our numerical examples, we will consider the potential parameterised as:
%%%
\be \label{eq:NumPot}
V(\phi)= \Lambda +\frac{1}{2}\mu \phi^2 +\frac{1}{3}\beta_3 \phi^3+\frac{1}{4} \beta_4 \phi^4 +  A \e^{-\alpha \phi^2}  \kma
\ee
%%%
We have chosen the following values for the constants appearing in $S_E$,
%%%
$
\kappa=0.1 \kma \; \xi=3 \kma \; \Lambda = 0.1 \kma \; \mu=1.0 \kma \;
\beta_3= -0.25 \kma \;  \beta_4 = 0.1 \kma \; A=3.0 \kma \; \alpha=2.0 \pkt
$
%%%
\begin{figure}[ht]
\centering
\includegraphics[width=\fullWidth]{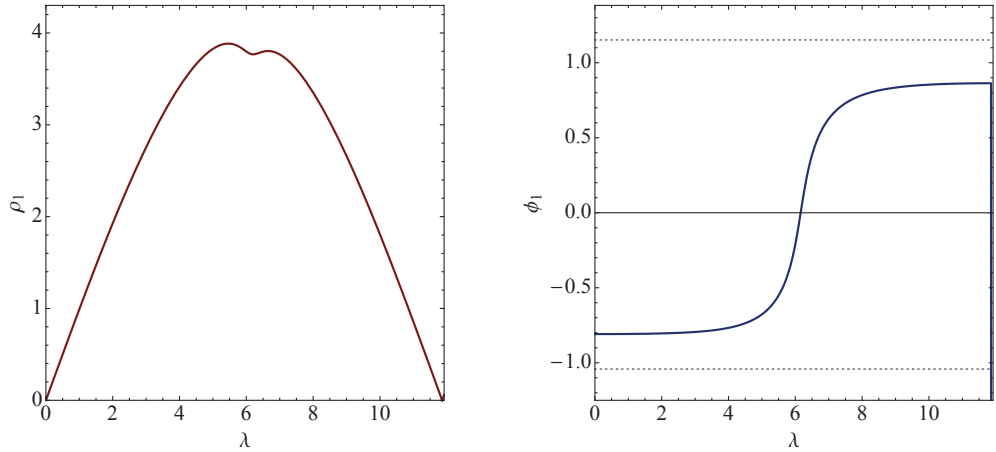} \caption{
\label{fig:fields1}
\small The field profiles (scale factor on the left, scalar field on the right) for our example of an instanton with a neck.}
\end{figure}
\begin{figure}[ht]
\centering
\includegraphics[width=\fullWidth]{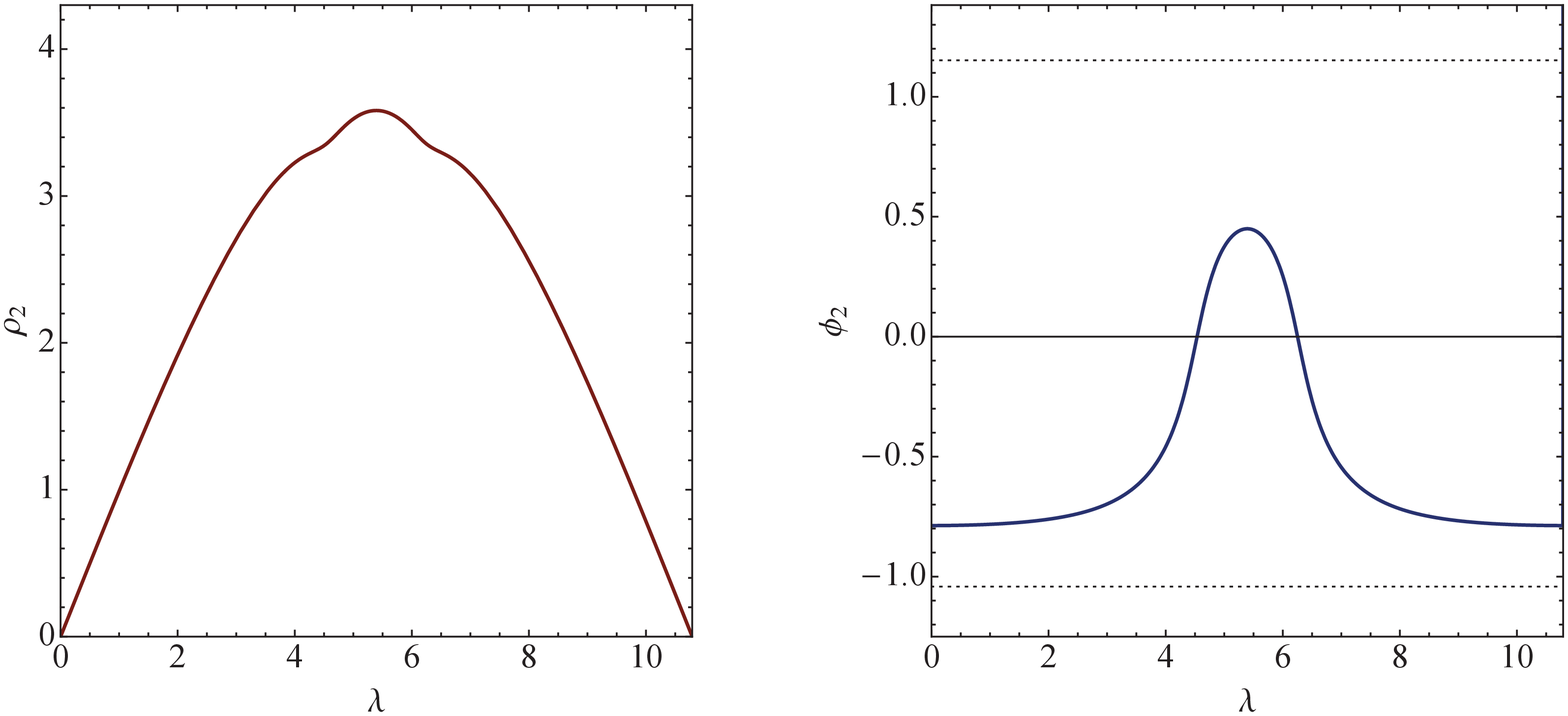} \caption{
\label{fig:fields2}
\small The field profiles (scale factor on the left, scalar field on the right) for our oscillating instanton example. The scalar field profile now leads to a hump in the scale factor, rather than a neck.}
\end{figure}
We have integrated Eqs.~(\ref{phi_dd2}) and (\ref{rho_2}) numerically with the boundary conditions
Eqs.~(\ref{eq:bcs}) and indeed found that instantons in this theory can have a neck.
An example of an instanton with neck is shown in Fig.~\ref{fig:fields1}. The scalar field has a characteristic kink profile while the scale factor $\rho$ develops a neck in the small $\phi$ region, where the suppression due to the factor $f^{-1/2}$ in Eq. (\ref{eq:rhotransf}) is the largest.
Note that in this potential one can also find oscillating instantons 
\cite{Hackworth:2004xb,Lavrelashvili:2006cv,Lee:2011ms,Battarra:2012vu,Battarra:2013rba}, 
in which the scalar field oscillates several times back and forth between the two sides of the potential barrier. 
An example of a twice oscillating instanton is shown in Fig.~\ref{fig:fields2}.
In this case the scalar field profile has two nodes and the scale factor acquires a ``hump'' instead of a neck. We should remark that, as already anticipated in \cite{gl87}, in order for these special features to arise the potential must contain a rather sharp barrier between the two local minima -- it is for this reason that we included a Gaussian term in our definition of the potential in Eq. (\ref{eq:NumPot}).
%%%%%%%%%%%%%%%%%%%%%%%%%%%%%%%%%%%%%%%%
\section{Concluding Remarks} \label{conclusions}
%%%%%%%%%%%%%%%%%%%%%%%%%%%%%%%%%%%%%%%%
We have shown that instantons with necks can be produced as a result of quantum tunnelling
in the decay of a metastable vacuum in scalar field theories with non-minimal coupling
to gravity (while they cannot be produced in the case of minimal coupling). After bubble materialisation, such neck geometries lead to two regions of the universe that are separated by a wormhole.
However, in contrast to the wormholes beloved by science fiction authors and as discussed in more detail in \cite{Battarra:2014naa}, these wormholes typically possess a wide cross-section.

It is important to stress that, while Jordan and Einstein frames are physically equivalent since they are simply related by a field redefinition, the choice of frame determines whether or not necks in the geometry exist. In fact it is the coupling of matter to gravity that determines what observers made of this matter will see. And in the example we have provided such observers would indeed see a wormhole. 
%%%
\section*{Acknowledgements}
\par
This work was supported by the ERC Starting Grant 256994 ``StringCosmOS''.
G.L. acknowledges support from the Shota Rustaveli NSF Grant No.~FR/143/6-350/14 
and Swiss NSF SCOPES Grant No.~IZ73Z0-152581.
%%%%%%%%%%%%%%%%%%%%%%%%%%%%%%%%%%%%%%%%
%%%

%%%
\end{document}